\documentclass[acmsmall]{acmart}
\usepackage{tabularx}
\usepackage{xcolor}
\usepackage{multirow}

\usepackage{listings}
\AtBeginDocument{%
  \providecommand\BibTeX{{%
    \normalfont B\kern-0.5em{\scshape i\kern-0.25em b}\kern-0.8em\TeX}}}

\setcopyright{cc}
\setcctype{by}
\acmJournal{PACMHCI}
\acmYear{2025} \acmVolume{9} \acmNumber{7} \acmArticle{CSCW332}

\acmMonth{11}  
\acmPrice{}
\acmDOI{10.1145/3757513}

\begin{document}

\title[Togedule: Scheduling Meetings with Large Language Models and Adaptive Representations \\ of Group Availability]{Togedule: Scheduling Meetings with Large Language Models and Adaptive Representations of Group Availability}

\author{Jaeyoon Song}
\affiliation{%
 \institution{Massachusetts Institute of Technology}
 \city{Cambridge}
 \state{MA}
 \country{United States}
}
\author{Zahra Ashktorab}
\affiliation{%
 \institution{IBM Research}
 \city{New York}
 \state{NY}
 \country{United States}
}
\author{Thomas W. Malone}
\affiliation{%
 \institution{Massachusetts Institute of Technology}
 \city{Cambridge}
 \state{MA}
 \country{United States}
}

\renewcommand{\shortauthors}{Jaeyoon Song, Zahra Ashktorab, and Thomas W. Malone}

\newcommand{\blue}[1]{{\textcolor{black}{#1}}}
\newcommand{\red}[1]{{\textcolor{black}{#1}}}

\received{July 2024}
\received[revised]{December 2024}
\received[accepted]{March 2025}

\begin{abstract}
\blue{Scheduling is a perennial—and often challenging—problem for many groups.} Existing tools are mostly static, showing an identical set of choices to everyone, regardless of the current status of \blue{attendees'} inputs and preferences. In this paper, we propose Togedule, an adaptive scheduling tool that \blue{uses large language models to} dynamically adjust the pool of choices and their presentation format. With the initial prototype, we conducted a formative study (N=10) and identified the potential benefits and risks of \blue{such an} adaptive scheduling tool. \blue{Then,} after enhancing the system, we conducted two controlled experiments, \blue{one} each for attendees and organizers \blue{(total N=66)}. \blue{For each experiment,} we compared scheduling with verbal \blue{messages, shared calendars, or} Togedule. Results show that Togedule \blue{significantly} reduces the cognitive load of attendees indicating their availability and improves the speed and quality of the decisions made by organizers.
\end{abstract}




\maketitle

\section{Introduction}

\begin{figure*}[h]
  \centering
  \includegraphics[width=\textwidth]{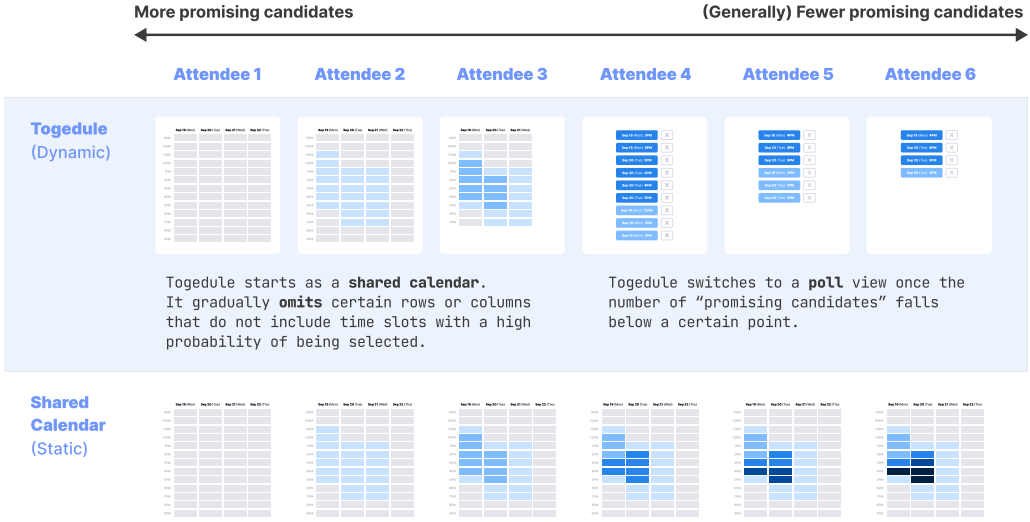}
  \caption{How \textsc{Togedule} adjusts the pool of choices based on prior inputs}
  \label{fig:dynamic}
\end{figure*}

It is often challenging to gather group availability and decide on the time to meet. Attendees often feel overwhelmed by the coordination efforts required for scheduling group meetings, and organizers wrestle with the complexity of processing attendees' responses.~\cite{romano2001meeting, mok2023challenging}.


To deal with this problem, some people hire administrative assistants who can schedule meetings on \blue{their} behalf. However, hiring an assistant is expensive, which is why most people still \blue{schedule using words in emails}~\cite{ducheneaut2001mail}. Nevertheless, scheduling meetings through verbal communication is often inefficient. For example, since \blue{only} limited information can be presented in a single email, there tend to be many back-and-forth emails asking for additional information. Moreover, it is \blue{often} burdensome for organizers to process long \blue{messages} to manually extract the availability information.

Existing scheduling tools such as shared calendars (e.g., When2meet) or polls (e.g., Doodle) provide a more efficient way of collecting and indicating the availability information. These tools, however, are static, meaning that the same full set of choices is presented to every attendee. Both shared calendars and polls show attendees all possible options even if some of them already have a very low probability of being accepted. \blue{But, they are best-suited} for different use cases: polls are more useful when there are \blue{relatively few} voting options, whereas shared calendars are more useful when there are \blue{relatively few} attendees.


In this paper, we propose the idea of an \textit{adaptive} representation of group availability that dynamically adjusts the pool and presentation of scheduling options using large language models (LLMs) (See Figure~\ref{fig:dynamic}).

To investigate the potential benefits and risks of adaptively representing the group availability, we designed the initial prototype and conducted a formative study with 10 participants. We found that with static tools attendees tend to indicate their availability partially. When there were only a small number of promising options shown, the attendees were more likely to mark not only options that definitely work but also options that might work. On the other hand, in static tools, since all options were shown no matter what, the attendee often assumed a certain set of options that seemed promising and omitted entering information about other options. They were also more likely to omit responses for time slots with intermediate popularity~\cite{zou2015strategic}. These behaviors can be problematic, especially when the attendee’s assumptions about the promising options turn out to be wrong. 

Based on these findings, we designed \textsc{Togedule}, a dynamic scheduling tool that continuously adapts to the inputs of attendees using an LLM. A \textit{dynamic scheduling tool} refers to a voting mechanism that updates the set of choices to be shown and the format in which they are presented in real time. For attendees, \textsc{Togedule} keeps adjusting the pool of candidate times based on how promising each time is and optimizes the presentation of these candidate times. By switching between poll and calendar views depending on the number of promising candidates, \textsc{Togedule} leverages the pros and cons of each format. In other words, it chooses the optimal variables as the x-axis and y-axis for each attendee. We used GPT-4 to keep the adaptation flexible and deal with various circumstances. For organizers, \textsc{Togedule} provides recommendations on which times to accept as the final meeting time taking into account the priority and preference of each attendee.

To evaluate the effect of \textsc{Togedule} in supporting attendees indicating availability and organizers deciding on the time to meet, we conducted two controlled experiments, one for attendees (N=48) and another for organizers (N=18). In both studies, we compared three conditions: scheduling in \blue{words (similar to email), a shared calendar, or \textsc{Togedule}}. Results show that attendees using \textsc{Togedule} had significantly less need for verbal communication when compared to using words only. Furthermore, attendees experienced significantly less mental load when using \textsc{Togedule} than using words or a basic shared calendar. For organizers, \textsc{Togedule} was found to improve the speed and quality of the decision on when to meet. Together, we believe our approach can help reduce the task load in the process of gathering group availability and increase the possibility of finding the time that works for more people in the group.

We do not propose our approach as a perfect solution for scheduling meetings. Our primary aim is to introduce the \textit{concept} of adaptive representation of group availability and present the initial evaluation results. We also do not argue that our approach is superior to all other existing tools. Rather, the idea of \textsc{Togedule} is complementary to and can be integrated with other solutions such as AI scheduling chatbots. Our idea advances the voting procedure among various components that constitute the scheduling process. In summary, this paper makes the following contributions. 

\begin{itemize}
\item \textsc{Togedule}, a novel scheduling tool that dynamically adjusts the pool and presentation of options using LLMs, based on the inputs of the attendees.
\item Results from a controlled experiment showing that \textsc{Togedule} helped attendees experience less cognitive load and enhanced the speed and quality of the decisions made by the organizers.
\item Findings from the formative study that revealed potential benefits and risks of dynamic scheduling tools
\end{itemize}

\section{Related Work}

\subsection{Scheduling Assistants and Verbal Communication}

To support the process of scheduling group meetings, some people hire assistants who can schedule meetings on \blue{their behalf}~\cite{ehrlich1986social}, which was found to help improve productivity~\cite{birkinshaw2013make}. In fact, most of the work done by administrative assistants involves scheduling work~\cite{erickson2008assistance}. However, hiring an assistant is expensive, which is why some people instead rely on software tools such as the ones elaborated in the following sections. Still, despite the presence of numerous scheduling tools, a survey reported that 80\% of information workers use email for scheduling~\cite{ducheneaut2001mail}. People prefer to use words in practice for various reasons. Existing tools may be burdensome or may not incorporate the subtleties that can be calibrated via words. There is also a social norm that people asking others to use a specific tool may seem bossy~\cite{nelson2008associations}.

Nevertheless, relying solely on verbal communication for scheduling meetings can prove highly inefficient~\cite{ducheneaut2001mail}. Processing numerous \blue{messages} at once through text can be challenging. For instance, the number of choices that can be effectively handled in a single email is limited, often resulting in multiple email threads going back and forth until an acceptable meeting time is agreed upon and details are finalized. Furthermore, organizers face significant time and effort in manually extracting availability information from long texts. These limitations led people to use shared calendars and polls to complement verbal communication, as described in the next section.

\subsection{Shared calendars and polls}

The most common type of software tools that support the voting process are shared polls and calendars. The shared polling approach lets organizers create polls with several date and time options and ask attendees to indicate their availability on the poll. Doodle is a representative example of a polling tool used worldwide and was found to successfully mitigate the burden of coordinating meetings~\cite{reinecke2013doodle}. In a typical shared poll, the column refers to a specific time and the row refers to each attendee.

Shared calendars are a graphical scheduling system where a user can drag to select dates they are available at~\cite{beard1990visual}. Typically, they show dates on the x-axis and times on the y-axis, which is the core difference from the polls. For example, When2meet~\cite{when2meet} is a shared calendar service that color-codes each time slot based on its popularity. Scheduling with shared calendars was found to be faster and less error-prone than manual scheduling~\cite{beard1990visual}. 

\begin{figure*}[h]
  \centering
  \includegraphics[width=1.0\textwidth]{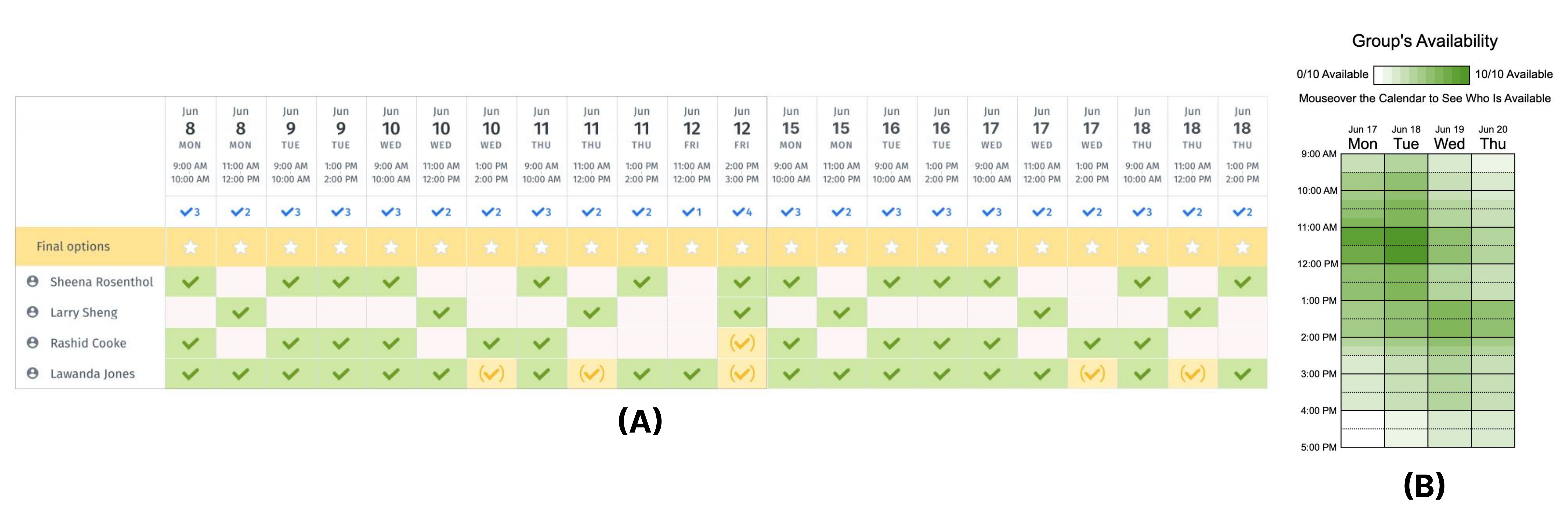}
  \caption{(A) A poll with many options is difficult to process (B) A shared calendar with many participants can be confusing. For instance, the difference between the color for ten votes and that for nine votes is so slight that it is hard to distinguish.}
  \label{fig:poll-limitation}
\end{figure*}

Although these tools are useful in scheduling meetings, both shared calendars and polls are static, meaning that they show an identical set of options to everyone from the start to the end, regardless of how likely each option will be accepted at the final time. Furthermore, polls are less useful when there are many voting options, as illustrated in Figure~\ref{fig:poll-limitation}. On the other hand, shared calendars can effectively represent many voting options, but are less useful when there are many attendees. In this paper, \textsc{Togedule} adjusts the pool of candidate times based on how promising each time is and renders the selected times as either a shared calendar or a poll. By switching between poll and calendar views depending on the number of promising candidates, we can leverage the pros and cons of each format. In other words, \textsc{Togedule} selects the optimal variables (e.g., time, date, attendee) as the x-axis and y-axis for each attendee.

\subsection{Finding a Mutually Agreeable Time}

Prior work has investigated ways to effectively identify a mutually agreeable time and handle complexities such as uncertainty~\cite{bowler2022exploring}. Romero et al. ~\cite{romero2017influence} discuss how early respondents in the scheduling process produce an information cascade that ultimately limits the options for the following respondents. In other words, the availability of early respondents is critical for the group scheduling to be successful. They also found that it is more likely to find a mutually agreeable time when a small number of respondents with the lowest availability are hidden from the poll. The design of the \textsc{Togedule} system takes into account these findings. For instance, \textsc{Togedule} shows the full set of options to the early respondents so as not to lose any useful information about their availability. Furthermore, when there is no optimal time that works for all attendees, \textsc{Togedule} excludes either the attendees with the least priority, and then recalculates the promising options.

Another work that analyzed the Doodle dataset~\cite{zou2015strategic} reported that open polls tend to have higher response rates for very popular and very unpopular time slots compared to hidden polls, whereas time slots with intermediate popularity have similar response rates. In \textsc{Togedule}, dynamic scheduling gradually excludes unpopular time slots to minimize the burden of checking the unpromising slots. 

Zhang et al. ~\cite{zhang2018understanding} conducted a field study with a mobile application for group event scheduling and found a series of factors that are significant in group event planning and decision-making process, such as host preference, user preference, and social voting influence. Considering their finding that later voters often face difficulty not being able to change the current status significantly, \textsc{Togedule} gives later voters an option to view more options where they can view less promising options as well. 

\subsection{Automated scheduling assistants}

A rich body of literature has explored delegating tasks to automated assistants~\cite{higel2003towards, halasz1994dexter, mitchellmcdermott, myers2007intelligent, singh2021automated}, with some of them focusing on automation of scheduling meetings in particular~\cite{hu2005preference, zunino2009chronos}. Despite the usefulness of these approaches, there is yet no tool that fully automates the whole scheduling process since it is often difficult for agents to understand all the nuances and complexities of what people want.

Accordingly, researchers have explored a hybrid approach that complements automation with the help of humans in the loop~\cite{teevan2016productivity, cranshaw2017calendar, vichivanives2023towards, papachristou2023leveraging, jeromela2023voicing}. For instance, Calendar.help~\cite{cranshaw2017calendar} delegates most of the scheduling work to an automated conversational agent and uses human assistants as a fallback for unusual scenarios. 

Pallagani et al.~\cite{pallagani2024prospects} organized the usage of large language models (LLMs) in planning and scheduling tools into eight categories: language translation, plan generation, model construction, multi-agent planning, interactive planning, heuristics optimization, tool integration, and brain-inspired planning. The LLM usage in \textsc{Togedule} falls into the categories of language translation, which involves converting natural language into structured planning languages and formats or vice-versa, and heuristics optimization, which applies LLMs in optimizing planning processes.

The process of scheduling meetings typically comprises two primary stages. The first step is to identify potential meeting times, which involves defining a set of candidate options. This step may be supported by partial automation such as importing existing calendars. Platforms like Google Calendar, Outlook, and Microsoft Teams offer functionality that automates this first step by cross-referencing multiple calendars to generate compatible time slots. Subsequently, the second stage entails selecting the final meeting time through a voting process. Despite advancements in automation, such manual voting remains necessary to ensure human confirmation and accommodate individual preferences. Online calendars may lack comprehensive information, such as unrecorded appointments or undisclosed preferences, necessitating human participation to finalize the meeting time.

For complex meetings, particularly those with a large number of attendees, it is often challenging for automated chatbots to converge on a meeting time, leading to many back-and-forth messages~\cite{cranshaw2017calendar, reinecke2013doodle}. Scheduling software tools like shared calendars or polls are alternative communication mediums that complement automated approaches. In this paper, we focus on advancing the voting mechanism for scheduling group meetings, which corresponds to the second aspect mentioned above. We believe an adaptive representation of group availability can be a good complement to automated scheduling chatbots and enhance the overall scheduling process.

\newpage
\section{Initial Prototype}
\label{sec:prototype}

\begin{figure*}[h]
  \centering
  \includegraphics[width=0.7\textwidth]{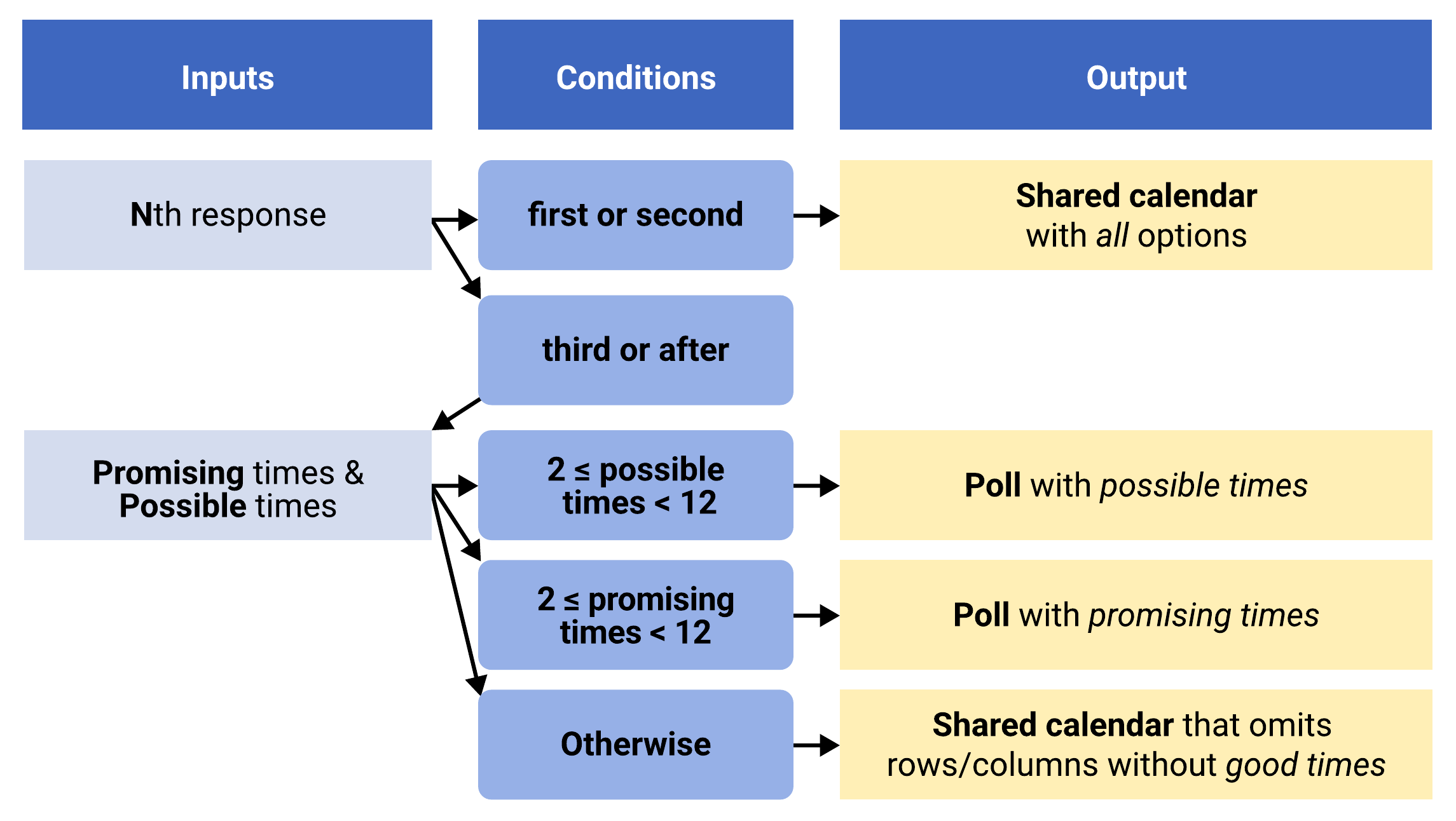}
  \caption{A diagram indicating how the dynamic adaptation works in the initial prototype}
  \label{fig:diagram}
\end{figure*}

\blue{\subsection{Key Challenges}}

\blue{Here, we summarize the key challenges and user needs that the prototype of \textsc{Togedule} aims to address. One primary challenge is reducing the dependence on verbal communication. While verbal exchanges can facilitate scheduling, excessive communication often leads to inefficiencies. Another challenge lies in leveraging the pros and cons of existing tools like polls and shared calendars, which rely on static designs that present fixed options to all participants. Polls with many options can overwhelm users, making it hard to identify the best choices. Shared calendars with many participants often become visually confusing. A further challenge involves the need for dynamic exclusion of unpromising options. Popular time slots are easily filled, but other slots require optimization to avoid burdening participants with unlikely choices. This suggests the need for dynamic scheduling mechanisms capable of intelligently adapting the representation of availability. Finally, organizers often face the burden of manually processing availability and prioritizing attendees.}

\blue{\subsection{Design}}

\blue{In response to these challenges, we designed an initial prototype that provides support for both attendees and organizers. The design process was led by the first author, drawing inspiration from existing applications such as polls (e.g., Doodle) and shared calendars (e.g., When2meet).} For attendees, the prototype dynamically reshapes the format in which the choices are presented. It shows a draggable calendar when there are many promising time blocks, and shows a poll when there are only a few promising candidates. For organizers, the prototype shows a table that summarizes the availability information, presents recommendations on the final meeting time, and provides priority settings where they can set the priority level of each attendee. The recommendations were automatically updated whenever organizers altered any of the priority settings.

Figure~\ref{fig:diagram} shows how we designed the dynamicity of the initial prototype for attendees. Specifically, the system first checks the number of inputs it has received to date. If the current attendee is the first or second person to enter the availability, the system shows the full set of candidates in a shared calendar format, unless there are fewer than twelve candidates in the initial set. This is because the poll view is not ideal for representing many candidates (See Figure~\ref{fig:poll-limitation}) and the availability of early respondents is vital for group scheduling to be successful~\cite{romero2017influence}. The first attendee has no choice but to view the shared calendar, whereas the second attendee has a choice to view fewer options, where they can start from only the times that work for the first person. From the third person, the system starts counting the number of ``promising'' and ``possible'' candidates. The ``promising'' candidates are the times that have the maximum number of attendees available. The ``possible'' candidates include all the ``promising'' times, along with the times that have one less attendee available. When calculating the ``possible'' candidates, the system first excludes either the attendees with the least priority or the ones with the lowest availability~\cite{romero2017influence}. If the number of ``possible'' times is less than twelve but more than two, the system shows a poll with ``possible'' times as candidates. If the number of ``promising'' times is less than twelve but more than two, the system shows a poll with the promising times. Otherwise, the system shows a shared calendar view but omits the rows and columns that do not include any of the ``good'' times. By saying \textit{good}, it means that more than 65\% of the attendees who already marked availability are available at that time. If the rows and columns are located in the middle of the calendar, not near the edge, they are not omitted to prevent confusion even if they do not contain ``good'' times.

We decided the threshold of two to twelve based on what is the number of times that can be well-represented in the poll form. If the attendee cannot see options at a glance, it means there are too many options to be represented. Furthermore, we chose to use 65\% to define the ``good'' times because we thought it was a high enough number to indicate that the option should not be omitted. We admit that these decisions were arbitrary and subjective without being fully justified, but they still successfully serve the purpose of testing an initial prototype. We further discuss these thresholds in section~\ref{sec:flexible}.
\section{Formative Study}

To educate our design of the dynamic scheduling tool, we conducted a formative study with the initial prototype. We sought to (1) investigate the practices and challenges of indicating availability and deciding on the time to meet using existing tools, and (2) identify potential benefits and risks of adaptively representing the group availability by testing out the prototype.

\subsection{Participants}
We recruited 10 participants (Male: 6, Female 4) on a freelancer hiring platform, Upwork. We divided \blue{the} 10 participants into two groups. Six participants were assigned the role of attendee, who indicated their availability and preferences. The other four participants were assigned the role of organizer, who decides on what time to meet based on the gathered information about the availability of the attendees. For the remainder of the paper, we refer to the former participants as \textit{attendees} and the latter participants as \textit{organizers}. This is to avoid confusion regarding study participants, which include both organizers and attendees.

On Upwork, we posted a job post that described the task. The post required the participants to be fluent in English. Freelancers satisfying this criterion applied for the job, and we selected the participants referring to their job success rate on Upwork. In the selection process, we varied the expertise of the participants such as data analysis, market research, web development, creative writing, academic writing, and email support. Each participant received 10 USD for completing a 30-minute study session.

In our job post for attendees, there were three tasks listed: selecting available times, responding to a survey, and possibly participating in a post hoc virtual meeting. We advised the attendees to think of the first two tasks as the process of scheduling the third task that may or may not happen. We clarified that this is an assumption for experimental purposes. \blue{The intention behind this design choice was to simulate a realistic scheduling scenario, encouraging participants to engage with the system as if they were coordinating an actual meeting.} We mentioned that they will be paid 10 USD for the first two tasks which will take about 30 minutes, and if they get to do the post hoc meeting, they will be paid a bonus payment. In reality, we ended up meeting with none of the attendees since we gathered enough data from the survey and via Upwork messenger.

\subsection{Tasks and Procedures}
We asked attendees to first watch a short video that shows how to use our prototype. We then asked them to indicate their schedule using (1) a shared calendar, (2) a shared poll, and (3) the prototype. The initial set of options were times between 9 AM and 5 PM on four weekdays. 

For all attendees, these options were within 1 month from the point that they performed the task so that they would reasonably know what their schedules looked like. \blue{We also informed them how important it is for them to attend the meeting, randomly assigning half of them as being ``very important'' and the other half as being ``less important.'' For instance, attendees assigned to the ``very important'' group were instructed to approach the meeting with a high priority, potentially being willing to adjust or reschedule other commitments to ensure their availability. Conversely, those assigned to the ``less important'' group were informed that their attendance was less critical and that they might prioritize other obligations over this meeting. This distinction aimed to mimic realistic variations in attendee priorities.} After they marked their availability, they responded to a survey asking about their scheduling practices and use of each system. The attendees were asked to describe how they performed the task, reflect on the strengths and weaknesses of the prototype, and comment on additional support they needed. 

For organizers, we first asked them to assume that they are a scheduling assistant and their job is to schedule a meeting with 6 potential attendees. We informed the organizers that one of the six attendees is particularly important to the meeting because they are the leader of the group. Then we advised them to use our prototype to browse the time preferences of attendees and decide on the best time to schedule the meeting. After making the decision, they answered a survey asking about the rationale for their decision, how they used each system, and aspects they liked and disliked about the prototype. We also asked follow-up questions to both attendees and organizers after the study via Upwork messenger.

\subsection{Findings}

Here, we summarize the main findings from the formative experiment and the survey.

\subsubsection{The need for a dynamic scheduling tool}

We found that the static tools that always showed the full set of options can overwhelm the attendees. This caused the attendees to arbitrarily omit to enter some of their availability information. One of the attendees mentioned that \textit{“There were just too many time slots, so I just checked the ones that seem popular.”} Four out of six attendees specifically mentioned that others’ availability matters. As these quotes suggest, when using static tools, attendees often assume a certain set of options that seem promising and omit entering information about other options. They were also more likely to omit responses for time slots with intermediate popularity~\cite{zou2015strategic}. There were times when the attendee’s assumption about the promising options was actually inaccurate --- some attendees failed to consider constraints like priority and preference of other attendees. For example, an attendee may omit their answers for an option they thought was not promising, but that option happened to be the most promising one after a few more responses. This suggests a need for a system to calibrate which options are promising on behalf of attendees to prevent inaccuracy. On the other hand, with a dynamic scheduling prototype, attendees were more likely to enter full information at least for the choices shown. They expressed that \textit{“When there were fewer times given, I also used the maybe mode.”}

Furthermore, the polling approach was found to be inappropriate for circumstances where there were many candidates to begin with (See Figure~\ref{fig:poll-limitation}). In these circumstances, shared calendars were more suitable to use.


Such limitations caused participants to end up relying on verbal communication again. For instance, one of the attendees tried to explain their availability information in an open-ended question of the survey since they felt it was not enough to mark it on the poll.

\subsubsection{Most attendees focus on the popular options}

We observed that most attendees do not consider the full set of candidate times in any case. Instead of selecting all times that work for sure and all times that might work, attendees tend to select only some of the possible times anyway. For example, one of the attendees mentioned in the survey response that \textit{“there exist other times that might have worked, but I did not select them because I wanted to emphasize what will work especially for me.”}

Then which set of candidates do the attendees consider? We found that other attendees’ choices influenced their selection. One of the attendees explained that \textit{“I just began from the times that had at least one person available since there were too many times to consider otherwise.”} Four out of six attendees specifically mentioned that others’ availability matters. They expressed that \textit{“then I clicked to see the different views … to see what I chose and how it lines up with other people's choices,”} and \textit{“I like seeing my choices reflected in real-time so I can see how they line up with other people’s selections.}

These findings suggest that attendees tend to narrow down the candidates by themselves, even if the system shows the full set of time choices. We do not mean that such behaviors are problematic. Rather, we think these behaviors are natural, and since attendees tend to indicate availability on only the promising ones anyway, our idea is that the system can process the promising options on behalf of users to make their jobs easier. In other words, it might not be necessary to show the full set of choices to everyone since users rarely review them all in any case. Showing only the selected subset of time choices that are promising can possibly reduce the mistakes of attendees in manually calibrating the promising candidates. It also helps decrease the time needed to mark availability, which aligns with Hick’s Law saying that the time taken to make a decision increases with the number of choices \cite{proctor2018hick}.

\subsubsection{Adaptive interface can cause confusion}

We found that attendees can find the dynamic change of the interface confusing. For instance, one of the attendees reported that \textit{“it was confusing because there was a difference when I came back to use the system after watching the video.”} This implies that the adaptive tool has a risk of confusion. Another attendee said, \textit{“lack of instruction and guidelines might be challenging,”} suggesting the need for letting the users know that the system is consistently adjusting the set of time options presented.

These findings resemble the guidelines for Human-AI Interaction that the system should update and adapt cautiously (G14) and inform users about changes (G18) \cite{amershi2019guidelines}. In this paper, to mitigate such confusion, we make clear on what basis the adjustment is taking place in the system. Furthermore, we designed the system to provide an option to bring back the unchanged system (i.e., show the initial full set of time blocks when requested) to prevent users from feeling that the change was disruptive.

\subsubsection{Complexities should be handled in an open-ended way}
\label{sec:flexible}

We observed that organizers consider a wide variety of factors when deciding the time to meet. Three organizers mentioned attendance as the important factor that affects their decision. They said, \textit{“I maximize the chances for everyone to come,”} \textit{“the most important one is the availability of each one,”} and \textit{“On 6/30, there is a chance that A, E, and/or F will show up thereby increasing participation”}. Two organizers mentioned that they prioritized important attendees. They reported that \textit{“on 6/27, the team lead will definitely be available and in most cases, the leader drives the rest of them. Unavailable team members can be easily updated afterward,”} and that \textit{“because at this time, F could also be there since is the project leader.”} System usage logs show that all four organizers used the priority settings in the prototype to prioritize the leader’s attendance. Other factors included timezone, urgency of the meeting, and the length of the window. Also, we observed that two organizers made assumptions about the probability of people showing up: \textit{“if they can or might be able to make 12 on 6/30, there may be other points during 6/30 they are available.”}

Considering such a wide variety, when scheduling tools try to incorporate various constraints, they should do it in an open-ended way because not all subtleties can be formalized. Findings indicate that it would be challenging to encode too many constraints at once. In other words, trying to deal with too many circumstances would often result in an information load. Therefore, in our design, we mostly focused on encoding the two most popular factors we found: attendance and priority. Rather than trying to provide a single recommendation that acts as an all-in-one solution, we propose providing multiple recommendations that assume different weights for each factor and leave the final decision to the human organizer.



In our prototype, we used arbitrary numbers as thresholds to decide how the system works for the purpose of initial testing. For instance, two and twelve are used as thresholds to decide the format of the availability information. 65\% was used to define the ``good'' times. However, using these fixed numbers can cause problems because they do not consider various constraints in different scheduling circumstances. It does not make sense to use the same threshold for different user groups regardless of their group size or the purpose of the meeting. For example, it may be more challenging to find a ``good'' time when there are more attendees in the meeting. Therefore, there is a need for a more flexible way to implement dynamicity in a way that copes with complex scheduling constraints. \blue{To address this, the following section introduces \textsc{Togedule}, a system that leverages the capabilities of large language models to make such complicated decisions with enhanced flexibility throughout the scheduling process.}




\section{\textsc{Togedule}: Adaptive Representation of Group Availability}

With these findings in mind, we designed \textsc{Togedule}, an adaptive scheduling tool that supports both attendees and organizers in scheduling a meeting. With \textsc{Togedule}, we envision attendees can be faster in selecting the times with less burden. Such an adaptive representation can also increase the possibility of finding the time that works for more people in the group by nudging the attendees to select popular times.

\begin{figure*}[h]
  \centering
  \includegraphics[width=\textwidth]{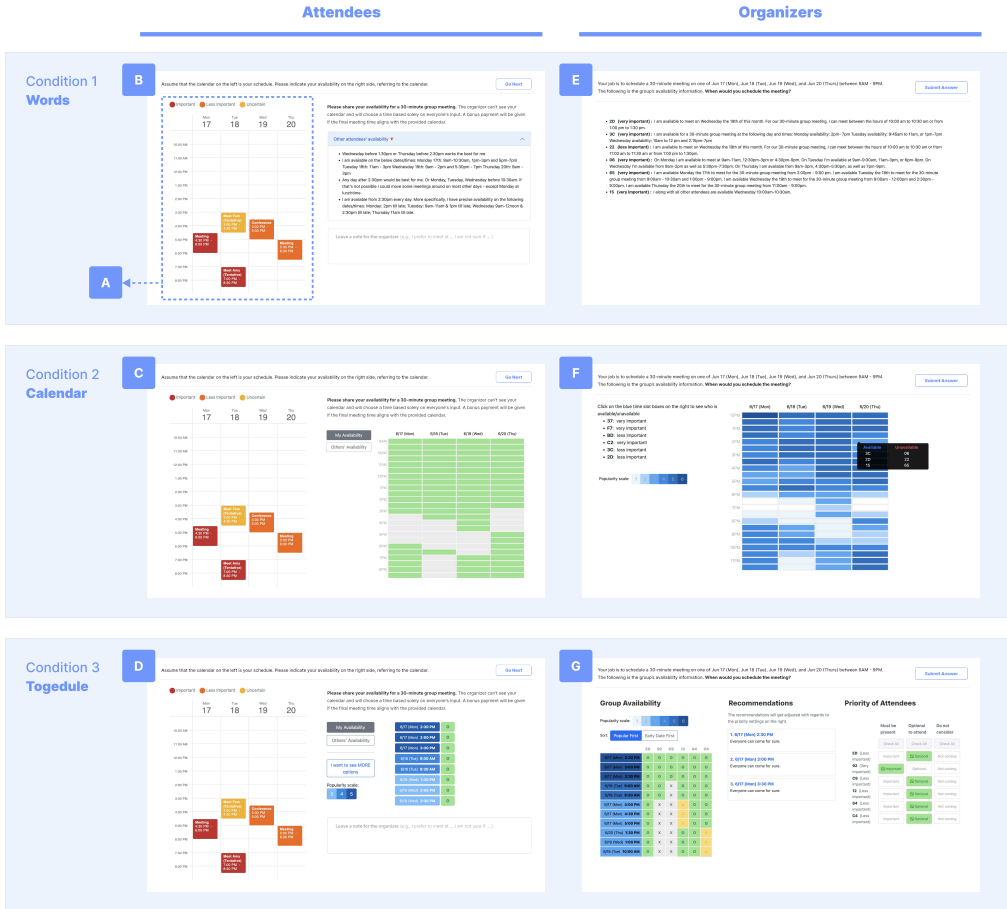}
  \caption{Interface for each condition: (A) calendar shown in all conditions for attendees to assume their schedule (B) the Words condition interface for attendees (C) the Calendar condition interface for attendees (D) the \textsc{Togedule} condition interface for attendees (E) the Words condition interface for organizers (F) the Calendar condition interface for organizers (G) the \textsc{Togedule} condition interface for organizers }
  \label{fig:conditions}
\end{figure*}

\textsc{Togedule} is a dual-sided system. The first side (Figure~\ref{fig:conditions}D) is designed to support attendees in marking their availability and preference information. The second component (Figure~\ref{fig:conditions}G) is designed to assist organizers in digesting the collected information and making the decision on when to schedule the meeting. 

\subsection{Support for Attendees Indicating Their Availability}

\subsubsection{A dynamic pool of choices}

The most novel aspect of \textsc{Togedule} is that it continuously adjusts the pool of candidate time slots presented to attendees based on how promising each candidate is. Initially, it evaluates the current votes to calibrate which time slots are likely to be accepted as the meeting time. This process identifies ``promising times,'' where the maximum number of respondents can attend, and ``possible times,'' accommodating the second largest group. With these calculations, \textsc{Togedule} uses GPT-4 to determine the degree of how many candidates to show. Finally, based on the decision of GPT-4, \textsc{Togedule} displays the selected time slots in the designated presentation format. \blue{Using GPT-4 allows \text{Togedule} to process complex inputs and dynamically balance competing factors, which makes it well suited to handle scheduling coordination in a flexible way. Of course, we considered developing a more mathematically based algorithm, but we believe it was more interesting to explore the performance of a modern LLM in making the decision.}

\begin{figure*}[h]
  \centering
  \includegraphics[width=\textwidth]{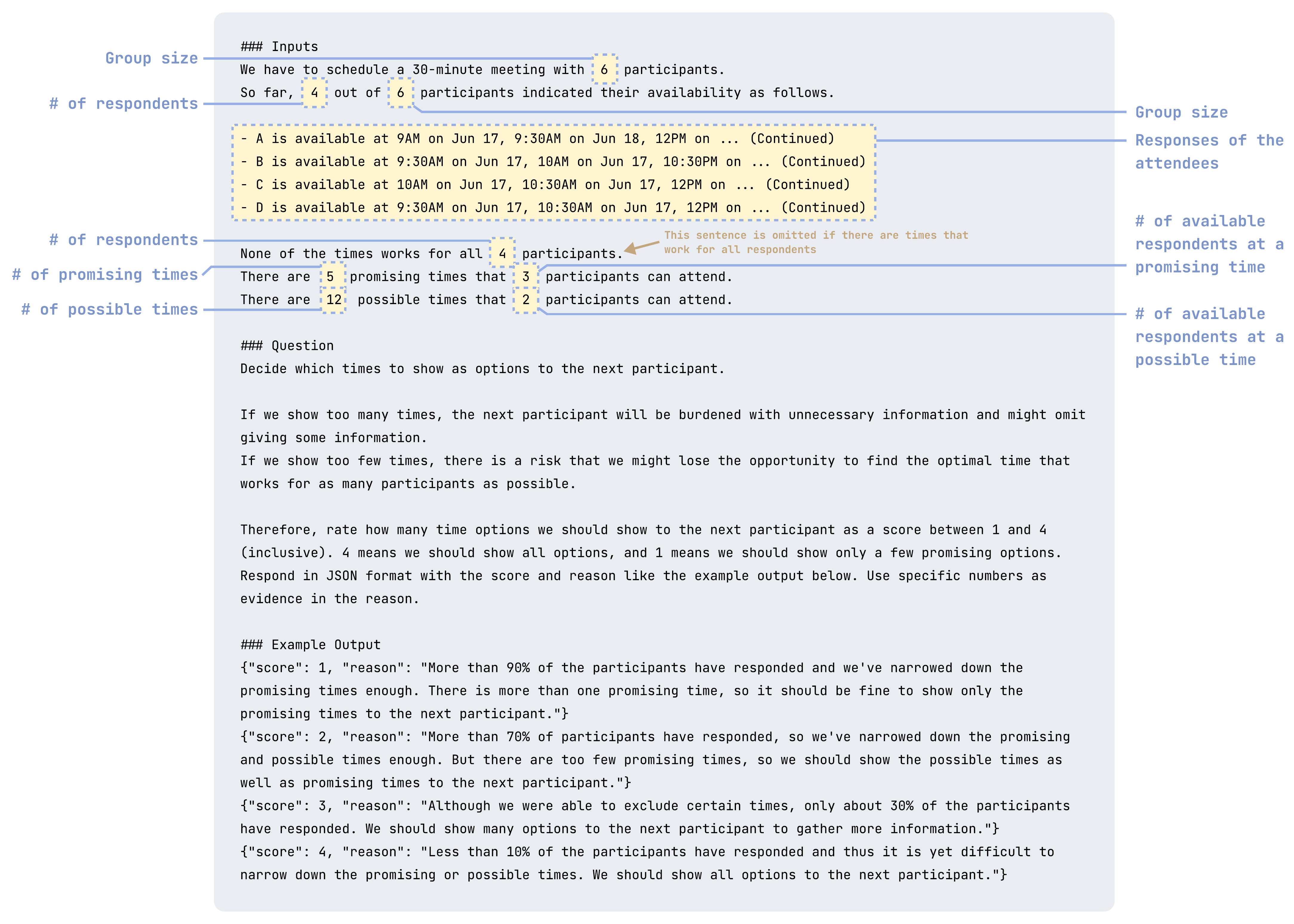}
  \caption{The prompt used to evaluate how many options to show}
  \label{fig:prompt}
\end{figure*}

Figure~\ref{fig:prompt} shows the prompt employed to elicit a rating from GPT-4 regarding the number of options to display. The prompt was finalized through iterative prompt engineering process~\cite{bsharat2023principled, clavie2023large, white2023prompt}. The colored text denotes variables contingent upon attendees' response status. \blue{For example, lines 5-8 represent the attendees' responses, which are generated based on their inputs provided through a poll or shared calendar and subsequently converted into text.} In cases where the promising times were identical to the options that work for all participants, the sixth line was excluded from the prompt.

In configuring GPT-4, we selected a temperature of 0.1, reflecting our assessment that the decision-making involved in this task leans towards being focused and deterministic rather than random. We still maintain that using GPT-4 enhances flexibility in that it can address diverse scenarios. It is just that each scenario is distinctly mapped to a particular decision outcome, which renders the decision-making in this task closer to determinism.

We opted to leverage GPT-4 to assess the voting status and provide a score, rather than tasking it with making detailed rendering decisions. This design choice stemmed from our prompt engineering efforts, which revealed that GPT-4 excels in generating a rating rather than handling every vote to deal with specific time slots. Additionally, entrusting GPT-4 with complete decision-making authority from start to finish appeared as a black box and was prone to randomness without sufficient justification.

Once the GPT-4 API returns a score and reason, \textsc{Togedule} accordingly renders its voting interface. If the score is 4, a shared calendar (Fig~\ref{fig:conditions}C) is displayed, showing all available options. In the case of a score of 3, \textsc{Togedule} presents a shared calendar with selective rows and/or columns omitted. To determine which rows/columns to omit, we leverage a rule-based algorithm. An option is deemed unpromising if the number of available attendees falls below half of the total attendees who have completed voting, provided that the ``promising times'' have more than half of the voters available. We arrived at this rule through an iterative design process to avoid excessive omission. If all options within a row/column are unpromising, it is omitted entirely. In the event of a score of 2, \textsc{Togedule} switches to a poll interface (Fig~\ref{fig:conditions}D) displaying ``possible times.'' For a score of 1, the interface shows a poll with ``promising times.'' This post-processing helps gradually change the presentation of options, as illustrated in Figure~\ref{fig:dynamic}.

In this way, our approach optimizes the number of choices presented to each attendee. Because this allows the attendees to review fewer options, \textsc{Togedule} can possibly help reduce their task load and improve their speed~\cite{proctor2018hick}. Since the presented subset of times is chosen in terms of popularity, this nudges the following attendees to select popular times, leading to a higher probability of converging.

Considering the finding from our formative study that dynamic scheduling tools can be confusing, \textsc{Togedule} provides a feature of seeing more options when attendees are shown a poll. This gives attendees the choice of whether to view the full set of options or the selected subset, thereby making the adaptation less disruptive. This also allows important attendees with a high priority to give more information about their availability.

\subsubsection{Multiple levels of preferences}

\textsc{Togedule} provides attendees with the flexibility to express three levels of preferences: available for sure, maybe available, and unavailable. By encoding such nuanced preferences, \textsc{Togedule} reduces the need for verbal communication. For instance, without this feature, attendees might have resorted to leaving notes saying that they did not select a certain time although it might work. Admittedly, there is a risk that incorporating such complexity may increase the task load for attendees, particularly if they perceive that too many things are represented in the system. However, we believe that improvement in the quality of the decision with the help of additional information would offset such a risk.

\subsection{Support for Organizers Deciding on the Time}

\subsubsection{Priority Management}

\textsc{Togedule} allows organizers to set up the priority of each attendee. They can select one of ``must be present,'' ``optional to attend,'' and ``not coming.'' If ``not coming'' is selected, their availability is no longer considered in the recommendation. Every time organizers change the priority information, \textsc{Togedule} accordingly updates the recommendations. 

\subsubsection{Multiple Recommendations}

There exists a degree of uncertainty surrounding complex constraints such as preference and priority in scheduling decisions. First, it is not clear how much weight we should put on times that \textit{might} work, in comparison to times that work for sure. Some organizers may decide that having two attendees who might attend is better than having one attendee attend for sure, while others may hold a different perspective. To address this issue, we employ two recommendation algorithms--one with a higher weight assigned to uncertain availability and another with a lower weight. Similarly, the prioritization of attendees presents a dilemma. For example, it remains unclear which to prioritize: one important attendee being available versus three optional attendees being available. To navigate this uncertainty, we utilize two algorithms--one that always prioritizes the important attendee and another that emphasizes overall attendance. By presenting multiple recommendations derived from different algorithms, we empower human organizers to make the final decision based on their unique considerations and preferences.

\subsection{Implementation}
We implemented \textsc{Togedule} as a JavaScript-based web app, and we
used an iterative design process of prototyping and pilot studies. \textsc{Togedule} is built with ReactJS on the front end and Express on the back end with MongoDB database. We also used the GPT-4 API provided by Open AI. \blue{The specific model used is ``gpt-4-turbo-2024-04-09.''} The code for \textsc{Togedule} is available as open-source software, along with the data collected from experiments described below (see \href{https://github.com/jyoonsong/togedule}{https://github.com/jyoonsong/togedule})



\section{Evaluation}

To assess the effects of \textsc{Togedule}, we conducted two controlled experiments: one for attendees indicating their availability (Study 1) and another for organizers deciding on the time to meet (Study 2).

\subsection{Conditions}
For both studies, we compared the Words (using words only), Calendar (using the basic shared calendar along with words), and \textsc{Togedule} (using \textsc{Togedule} in addition to words) conditions. The experimental interface used in each condition is shown in Figure~\ref{fig:conditions}. 

For two primary reasons, we chose verbal communication and basic shared calendars as baselines for comparing \textsc{Togedule}. Firstly, they are \blue{two of the most} widely adopted methods for scheduling group meetings. Verbal communication via emails remains a popular mode of interaction for scheduling meetings~\cite{ducheneaut2001mail}. Additionally, shared calendars such as When2meet are prevalent tools for coordinating group schedules. Secondly, each baseline method offers clear advantages. Verbal communication boasts high accessibility and shared calendars excel in managing a large number of options, overcoming the limitations often associated with alternative polling methods (refer to Figure~\ref{fig:poll-limitation}). Using external tools directly as baselines was not feasible due to the inability to record accurate metadata, such as speed. Additionally, we aimed to mitigate confounding factors arising from minor visual differences and supplementary features that are not central to our study. Thus, we took the initiative to implement the baseline interfaces on our own, replicating the functionalities of existing tools as much as possible.

\subsection{Participants}
We recruited a total of 66 participants (48 attendees and 18 organizers) through Prolific. To induce their behaviors to be representative of real-world users scheduling meetings, we devised an incentive scheme aimed at motivating users to accurately mark their availability based on provided artificial schedules. The base payment for participating in the experiment was 5 USD for attendees (20 minutes) and 4 USD for organizers (15 minutes). Attendees were informed of a bonus payment of 1 USD per condition if the final meeting time aligned with the artificial schedule provided, potentially totaling up to 3 USD in bonus payments. Organizers, on the other hand, were advised that their bonus payment, up to 3 USD, would be contingent on the quality of their decisions as rated by a professional scheduling assistant. We divided the attendees into 6 groups--three consisting of 6 attendees and three consisting of 10 attendees. We opted not to include smaller groups in our study, as \textsc{Togedule} mainly targets complex scenarios typically involving larger groups. \blue{We chose not to consider groups with more than 10 attendees because they may be relatively less common in typical meeting scenarios~\cite{cao2021large}}. Each organizer assessed two groups across three conditions, resulting in a total of 6 tasks per organizer.

\subsection{Tasks and Procedures}
Our experiments used a within-subject design, where each participant performed the task for all three conditions. \blue{We counterbalanced the order of conditions across participants. Given three conditions to compare, there are six possible orders in which these conditions can be presented. These six orders were distributed among 48 attendees and 18 organizers, resulting in 8 attendees and 3 organizers for each order. The attendees were randomly assigned to either a six-person group or a ten-person group. Thus, the first and second attendees in each group experienced different orders of conditions. This prevents potential bias arising from the influence of early respondents. The studies were approved by the IRB of our institution.}

In Study 1, attendees were given a calendar populated with either 5 or 10 events, randomly distributed from 9 AM to 9 PM across four days (Fig~\ref{fig:conditions}A). We varied the number of events to simulate differing levels of busyness: some calendars were densely packed with 10 events, while others were relatively less busy with only 5 events. Additionally, we randomly assigned priority to each event, some as tentative and some as particularly important. Furthermore, we varied the time preference of each attendee: some favored evenings, indicated by a higher density of existing events before 3 pm, while others preferred mornings, reflected by a greater concentration of existing events after 3 pm. Attendees were instructed to regard the provided calendar as their own and accordingly indicate their availability within the given interface. 

In the Words condition, attendees were given a text input where they could explain their availability and preference in natural language (Fig~\ref{fig:conditions}B). The Calendar condition provided a shared calendar where attendees can drag to select times that work for them (Fig~\ref{fig:conditions}C). Lastly, the \textsc{Togedule} condition asked the attendees to use \textsc{Togedule} in indicating their availability (Fig~\ref{fig:conditions}D). In the Calendar and \textsc{Togedule} conditions, attendees were able to optionally leave notes if needed. For each condition, attendees were randomly assigned whether it was more or less important for them to attend the meeting.

In Study 2, organizers were asked to assume that they were a scheduling assistant and decide on the time for designated attendees to have a 30-minute meeting. We advised them to refer to the information in the provided system when making the decision. First, in the Words condition, there was a list of texts the attendees left (Fig~\ref{fig:conditions}E). Second, the Calendar condition showed a shared calendar that color codes the popularity of each time block (Fig~\ref{fig:conditions}F). In the final condition, organizers used the \textsc{Togedule} system (Fig~\ref{fig:conditions}G). In all three conditions, organizers were given information about the priority of each attendee, whether they were vital or optional to be in the meeting. After using the provided system, organizers submitted the date and time they decided on.

At the beginning of the experiment, all participants answered a pre-survey that asked about demographics and their regular practices of scheduling meetings. After completing a task for each condition, participants answered a post-survey about their use of the system.  

\subsection{Measures and Analysis}
We recorded the time taken to investigate the effect of \textsc{Togedule} on speed. In addition, we measured participants’ task load and user experience for each condition with 5-point Likert scale questions. We used NASA-TLX to assess the task load and tweaked the Single Ease Question (SEQ) to investigate the ease of use. We also asked participants to describe the process of performing the task and aspects they liked or disliked about each condition.

For attendees, we additionally asked how useful each condition was in indicating their schedule, how much it helped express their preference between different times, and whether they tried to select popular times as much as possible. For organizers, we asked how useful each condition was to consider the priority of each attendee and how useful it was to consider the attendees’ preferences.

\begin{table*}
  \caption{The five labels we used to open-code the notes that attendees left}
  \label{tab:note_category}
  \begin{tabularx}{\textwidth}{|l|X|X|}
    \toprule
    Label & Description & Example \\
    \midrule
    Availability & Explanation on what times work for them & I'm available on the date of Monday 19 from 10 AM to 9:30 AM. \\
    Preference & Expression of preference for certain time blocks & I prefer to meet in the morning. \\
    Priority & Notes mentioning how important the meeting is to them & I am less important in this meeting. \\
    Reasons & Reasons why they indicated their availability in the way they did & I have to attend the interview and meeting which are held on Thursday. \\
    Information & Notes regarding other information & I prefer to meet at the office. \\
    \bottomrule
  \end{tabularx}
\end{table*}

To analyze how \textsc{Togedule} substitutes verbal communication, we conducted a discourse analysis with the notes that attendees left. We open-coded each note in terms of the purpose they serve. There were five labels: availability, preference, priority, reasons, and information. The definition of each label is described in Table~\ref{tab:note_category}. The notes irrelevant from scheduling (e.g., asking for a bonus payment) were excluded from the analysis.



\blue{We also analyzed the quality of decisions made by the organizers. We conducted an analysis using ratings provided by 16 Prolific raters. These raters were carefully selected based on the following criteria: they were native English speakers with an approval rate of 99\% or higher, had completed at least 200 tasks on Prolific, held a degree from a technical or community college or higher, and had not participated in our experiments as either attendees or organizers. The raters used a website we provided to assess each decision on a 7-point Likert scale. For their work, they were compensated 6 USD for rating 18 decisions. Each decision was evaluated by 5 or 6 raters. Decisions were classified as ``good'' or ``bad'' based on whether their ratings were higher than the average score. The reliability of these ratings was evaluated using the intraclass correlation coefficient (ICC-3k), yielding a value of 0.751, which indicates a fair level of agreement among raters.}

\section{Results}

\subsection{Overall system usage}

We report the overall usage of the \textsc{Togedule} system here. Out of 48 attendees, 25 attendees were initially shown a shared calendar view, and the other 23 attendees began from the poll view. The attendees switched from shared calendar to poll to view fewer options 62 times, and switched from poll to shared calendar to view more options 71 times in total. Furthermore, they checked the availability information of other attendees 286 times in total, which is 5.96 times per attendee on average. Among the 54 decisions submitted by 18 organizers, there were 33 good decisions.

\subsection{Attendees}

\begin{figure*}[h]
  \centering
  \includegraphics[width=\textwidth]{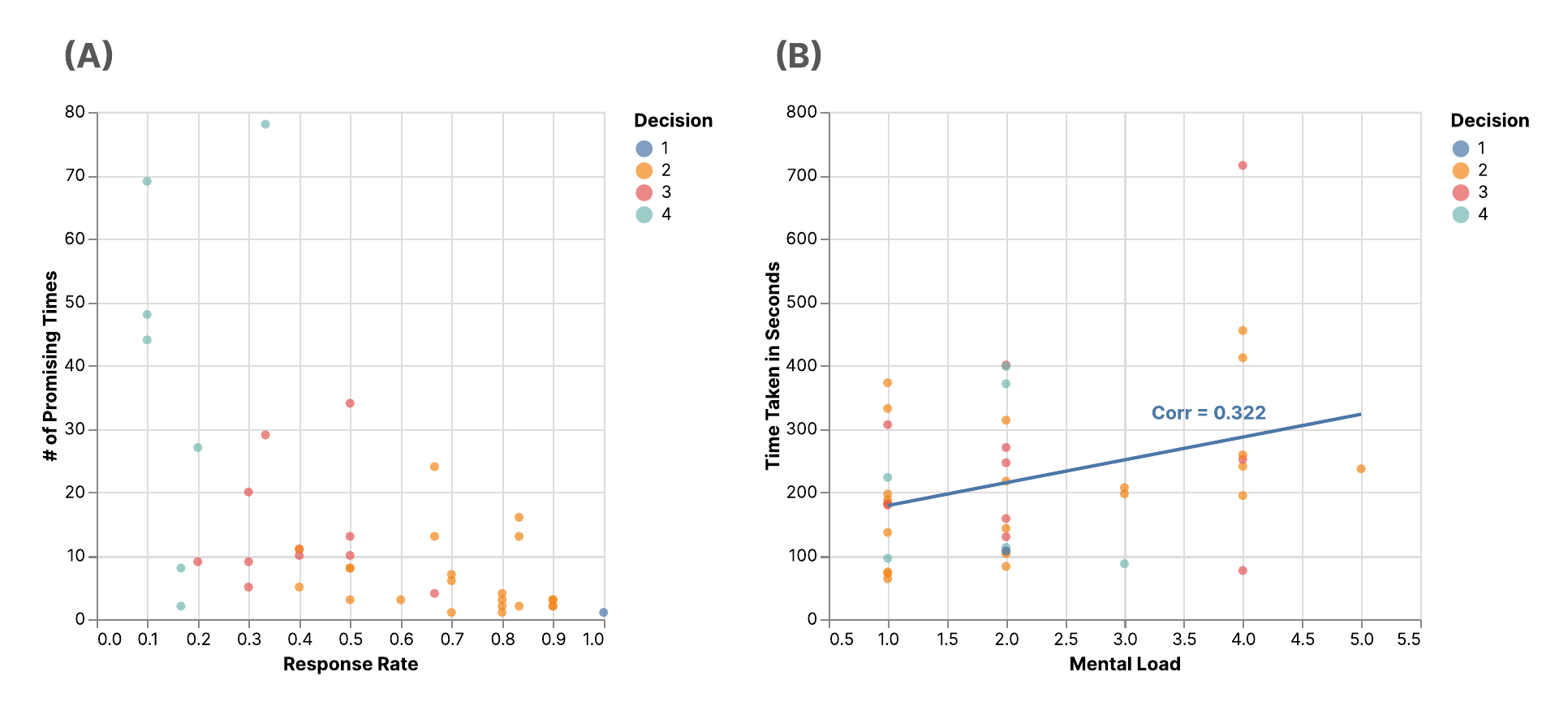}
  \caption{Decisions of GPT-4: (A) the relationship between the response rate and the number of promising times at the time when GPT-4 made the decision. (B) the relation between mental load and time taken in seconds using \textsc{Togedule} in the task. In both graphs, the color denotes the final decision made by GPT-4.}
  \label{fig:llm-decisions}
\end{figure*}

\subsubsection{Overview of GPT-4 Decisions}

\newcolumntype{b}{X}
\newcolumntype{m}{>{\hsize=.35\hsize}X}
\newcolumntype{s}{>{\hsize=.2\hsize}X}

\begin{table*}
\fontsize{8pt}{8pt}\selectfont
  \caption{The examples of GPT-4's evaluation of how many time options should be shown to the next participant }
  \label{tab:gpt-examples}
  \begin{tabularx}{\textwidth}{s|m|m|m|s|b}
    \toprule
    Group Size & \# of Respondents (Response Rate) & \# of Promising Times & \# of Possible Times & Decision of GPT-4 & Reason Provided by GPT-4 \\
    \midrule
    10 & 10 (100.0\%) & 1 & 1 & 1 & All 10 participants have already indicated their availability, and we have identified 1 promising time where all participants can attend. Given that we have a perfect match for all participants, it is most efficient to show only this promising time to the next participant. \\
    \midrule
    10 & 5 (50.0\%) & 3 & 5 & 2 & Half of the participants have already provided their availability, allowing us to identify some promising and possible times. However, with only 3 promising times and 5 possible times identified, it's prudent to show both promising and possible times to the next participant to maximize the chances of finding a time that suits the majority. \\
    \midrule
    6 & 3 (50.0\%) & 34 & 95 & 3 & Only 50\% of the participants have responded so far, providing a limited view of the overall availability. With 34 times where 2 participants can attend and 95 times where only 1 participant can attend, it's important to show a broader range of options to the next participant to increase the chances of finding a time that suits more attendees. \\
    \midrule
    10 & 1 (10.0\%) & 48 & 48 & 4 & Only 10\% of the participants have responded so far, providing availability for 48 times. Given the low response rate, it is crucial to show all available options to the next participant to maximize the chances of finding a time that suits everyone. \\
    \bottomrule
  \end{tabularx}
\end{table*}


Table~\ref{tab:gpt-examples} shows the examples of how GPT-4 decided the number of options to be presented. GPT-4 generally gave a low score when the response rate was low and a high score when the response rate was high. Exceptionally, it gave a different score if the number of promising and possible times were particularly low or high even when the response rate is the same, as shown in the second and third examples in Table~\ref{tab:gpt-examples}. 

Figure~\ref{fig:llm-decisions} shows two scatter plots regarding the GPT-4 usage. The first plot shows the distribution of each score depending on the response rate and the number of ``promising times.'' We can observe that the decision score has a strong relation with both variables. The second plot visualizes the correlation between the mental load and the time taken in seconds for using \textsc{Togedule} in the task. The Pearson correlation coefficient between these two variables is shown to be 0.322 ($p < 0.05$), but neither of them shows a clear correlation with the decision score.

\begin{figure*}[h]
  \centering
  \includegraphics[width=\textwidth]{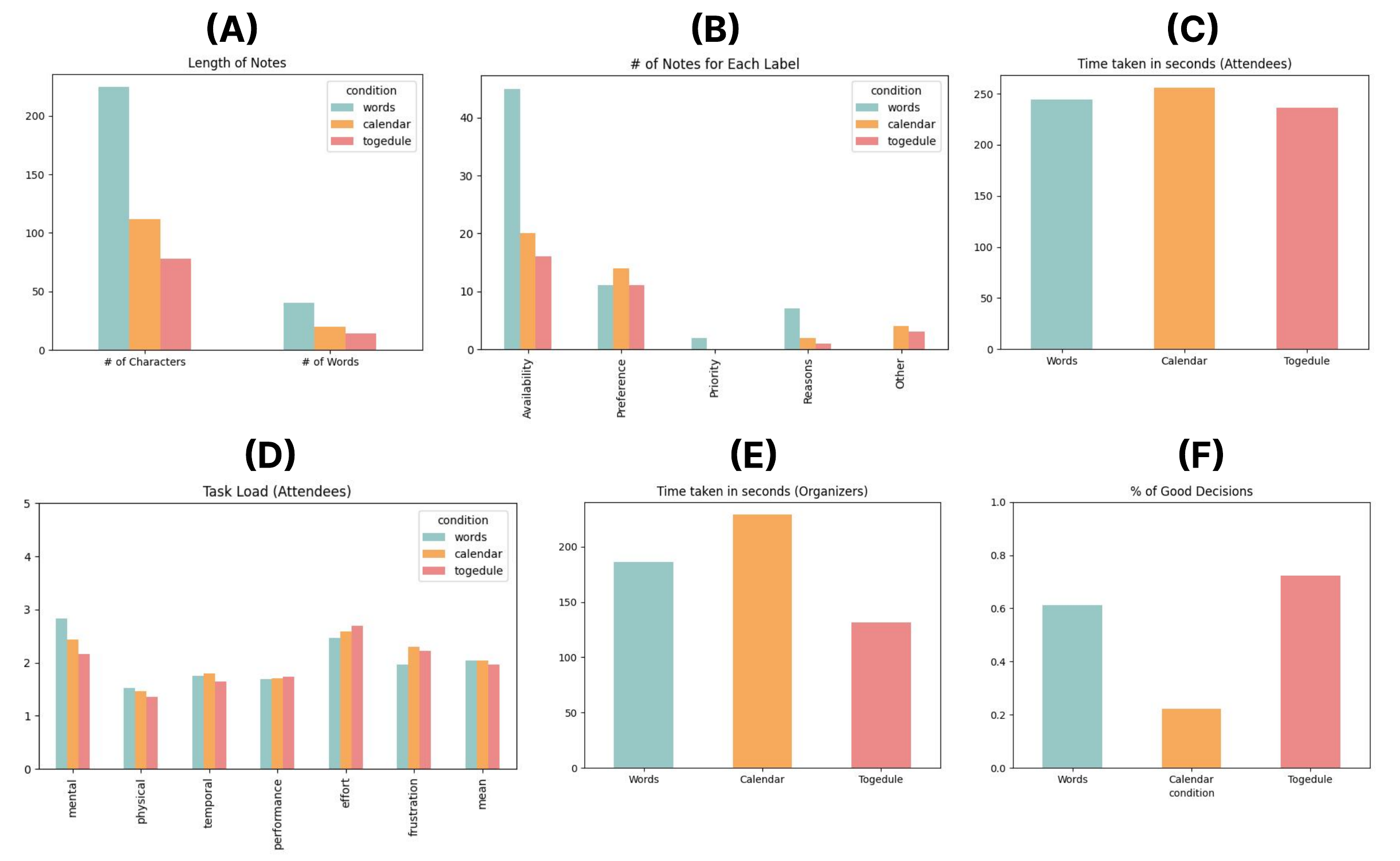}
  \caption{Results from the experiments: (A) the average length of notes in characters and words (B) the average number of notes for each label (C) the mean time taken in seconds for attendees (D) the mean task load indices for attendees (E) the mean time taken in seconds for organizers \blue{(F) the proportion of good decisions}}
  \label{results}
\end{figure*}




\subsubsection{Reduced verbal communication}
The notes attendees left across three conditions were 138.06 characters and 24.79 words long on average. The notes were the shortest in the \textsc{Togedule} condition on average as shown in Figure~\ref{results}A. We also ran a t-test on the number of words in the notes. In general, we found that \textsc{Togedule} statistically significantly reduced the number of words used compared to the Words condition ($p < 0.01$ and $t=7.245$). \red{The Cohen’s $d$ for this comparison was 1.046, indicating a large effect size.} The reduction in comparison to the Calendar condition was not statistically significant. This implies that \textsc{Togedule} reduces verbal communication compared to using words alone, but not compared to using shared calendars.





We also compared the number of words used for specific purposes using the five labels that we open-coded on the notes (See Figure~\ref{results}B). Notes regarding availability were significantly reduced in the \textsc{Togedule} condition when compared to the Words condition ($p < 0.01$ and $t = 8.470$). \red{The Cohen’s $d$ for this comparison was 1.223, indicating a large effect size.} Furthermore, \textsc{Togedule} significantly reduced the notes taken about the reasons for available or unavailable times compared to the Words condition ($p < 0.05$ and $t = 2.205$). \red{The Cohen’s $d$ for this comparison was 0.318, indicating a moderate effect size.} Other labels did not show any statistically significant difference between study conditions.





\subsubsection{Speed of indicating availability}
The average time taken in seconds for each condition is shown in Figure~\ref{results}C. The \textsc{Togedule} condition shows the lowest mean ($M=236.062, SD=155.355$). The Words condition is the next ($M=244.107, SD=168.614$), and the Calendar condition was the highest ($M=255.782, SD=185.822$). The differences between each condition were not statistically significant using a t-test. For example, there was no significant improvement using \textsc{Togedule} rather than Words ($p = 0.749$ and $t = 0.322$). \red{The Cohen’s $d$ for this comparison was 0.047, indicating a negligible effect size.} This suggests that speed is not the main area that \textsc{Togedule} enhances in the scheduling process. The median time taken is 196.876 seconds for the \textsc{Togedule} condition, 201.948 seconds for the Calendar condition, and 199.113 seconds for the Words condition.


\subsubsection{Task load of attendees}
The average task load indices are shown in Fig~\ref{results}D. The level of mental load, physical load, and temporal load were the lowest in the \textsc{Togedule} condition on average. Attendees experienced a lower level of frustration in the \textsc{Togedule} condition than in the Calendar condition on average, but not when compared to the Words condition. Only the mental load was shown to be significantly lower in the \textsc{Togedule} condition than in the Words condition ($p < 0.05$ and $t = 2.483$). \red{The Cohen’s $d$ for this comparison was 0.358, indicating a moderate effect size.} The difference between the Calendar and \textsc{Togedule} conditions was not statistically significant ($p = 0.170$ and $t = 1.391$). \red{The Cohen's $d$ for this comparison was 0.201, representing a small effect size.}

On the other hand, the perceived performance was slightly worse with \textsc{Togedule} than with the Calendar and Words conditions. The perceived effort level was also the highest in the \textsc{Togedule} condition. However, none of these comparisons were statistically significant.  

To further investigate the increase and decrease in the task load, we analyzed the open-ended answers from the post-survey. One of the respondents mentioned that \textsc{Togedule} was \textit{Easy to see things at a glance, easy to see what times were popular with others without having to give it much thought.}. Another respondent also said it was \textit{Very visual and easy to understand at a glance.} These imply that narrowing the pool of choices might have helped reduce the mental load of the task.


\subsubsection{Usefulness of \textsc{Togedule} in indicating availability}
64.583\% of attendees agreed or strongly agreed that \textsc{Togedule} was easy to use. 85.417\% of attendees agreed or strongly agreed that \textsc{Togedule} was useful in indicating their schedule. 68.75\% of attendees agreed or strongly agreed that they tried to choose the popular times as much as possible when using the system.

At the end of the experiment, we asked attendees which system they most preferred. 23 respondents answered they preferred \textsc{Togedule} the most, 17 respondents picked the Calendar condition, and 8 selected the Words condition.

\subsection{Organizer}




\subsubsection{Quality of the decision}

Overall, \textsc{Togedule} helped improve the quality of the decisions made by organizers (See Fig~\ref{results}F). 72.22\% of the decisions were considered good in the \textsc{Togedule} condition, whereas 22.22\% and 61.11\% were considered good in the Calendar condition and the Words condition, respectively. We found that \textsc{Togedule} enhanced the mean quality of decisions compared to both the Words condition and the Calendar condition. The difference between the \textsc{Togedule} and Calendar conditions was statistically significant ($p < 0.01$ and $t = 2.965$) \red{The Cohen's $d$ for this comparison was 0.699, indicating a moderate effect size.} However, the difference between the \textsc{Togedule} and Words conditions was not statistically significant.

We also performed a t-test on the binary scale, categorizing the decisions as good or bad depending on whether their rating was higher than the average rating, which was 4.598 out of 7. In this test, the quality of decisions made on \text{Togedule} was significantly better than that of the Calendar condition ($p < 0.01$ and $t = 3.0$), \red{with a Cohen's $d$ of 0.707, indicating a moderate effect size. However, the difference with the Words condition was not significant.}

\subsubsection{Speed of the decision}
The mean time taken in seconds for each condition is shown in Fig~\ref{results}E. We ran a t-test on the time taken to decide on the time to meet. Organizers were statistically significantly faster when using \textsc{Togedule} than the basic shared calendar in the Calendar condition ($p < 0.05$ and $t = -2.606$). \red{The Cohen's $d$ for this comparison was 0.614, indicating a moderate effect size.} Organizers were slightly faster when using \textsc{Togedule} than words as shown in Fig~\ref{results}E, but the improvement was not statistically significant. The additional cognitive load due to the information load in the \textsc{Togedule} system might have offset the improvement due to the support that \textsc{Togedule} provides.







\subsubsection{Task load of organizers}
On average, organizers reported significantly better self-perceived performance ($p < 0.05$ and $t = 2.250$) and less effort ($p < 0.05$ and $t = 2.695$) when using \textsc{Togedule} than the Calendar condition.  \red{The Cohen's $d$ for these comparisons were 0.530 and 0.635, respectively, indicating moderate effect sizes.} Organizers also showed significantly lower mental load ($p < 0.01$ and $t = 4.610$) and less effort ($p < 0.01$ and $t = 5.169$) with \textsc{Togedule} when compared to the Words condition. \red{The Cohen's $d$ for these comparisons were 1.087 and 1.218, respectively, both indicating large effect sizes.}






\subsubsection{Usefulness of \textsc{Togedule} in deciding the time to meet}
83.33\% of organizers agreed or strongly agreed that \textsc{Togedule} was easy to use. 88.89\% of organizers agreed or strongly agreed that \textsc{Togedule} was useful in deciding the time to meet. 83.33\% of organizers agreed or strongly agreed that \textsc{Togedule} helped them consider the priority of each attendee in making decisions. 88.89\% of organizers agreed or strongly agreed that \textsc{Togedule} helped them consider the preference of each attendee in making decisions. Organizers said \textsc{Togedule} was significantly more useful than using words on average using a t-test ($p < 0.05$ and $t = 2.538$). \red{The Cohen's $d$ for this comparison was 0.598, indicating a moderate effect size.} Organizers also thought \textsc{Togedule} was significantly more helpful in considering the priority ($p < 0.05$ and $t = 2.236$) and preference ($p < 0.01$ and $t = 3.059$) when compared to the Calendar condition. \red{The Cohen's $d$ for these comparisons were 0.527 and 0.721, respectively, both indicating moderate effect sizes.}

\section{Discussion}

\subsection{AI-assisted group scheduling}

\textsc{Togedule} can be a good complement to automated tools such as conversational agents. For instance, automated agents can manage the overall coordination process by sending out emails to attendees~\cite{cranshaw2017calendar}, but include the \textsc{Togedule} link in the emails. Then the attendees can mark their availability on \textsc{Togedule}, without the need for replying in words. \blue{Such integration can reduce the reliance on back-and-forth communication and processing long texts.}

It would be interesting to further extend the artificial intelligence component in \textsc{Togedule}. For example, instead of asking the attendees about their preferences every time they have a meeting to schedule, \textsc{Togedule} can learn their preferences over time and update them only when needed. \blue{The system could observe patterns in previous choices and adapt accordingly, updating these preferences only when significant deviations occur.} Future work can explore these opportunities for intelligent systems that learn scheduling practices, \blue{by measuring improvements in scheduling efficiency and accuracy in predicting availability}.

\blue{It is important to note that learning user preferences over time can introduce ethical concerns. While this capability can improve the user experience by streamlining scheduling and reducing repetitive inputs, it also risks exposing sensitive behavioral patterns. For example, the system might infer personal habits, such as preferred work hours or regular commitments, which could be misused if not properly protected. To address these concerns, the system should provide users with control over their learned preferences, including the ability to review, modify, or delete them. Expanding on the ethical considerations of using AI in scheduling, privacy and data handling are also critical. Scheduling tools often process sensitive user information, such as health-related data (e.g., doctor’s appointments) or confidential meetings, which raises important questions about data security and user consent. \textsc{Togedule} seeks to mitigate these concerns by allowing users to mark their availability without requiring them to disclose specific reasons for being available or unavailable, thereby minimizing the collection of sensitive data. Future work could build on this foundation by implementing additional safeguards, such as limiting data retention periods and offering transparency in how data is used.}

Furthermore, there is a need to improve the explainability of how the LLM makes decisions in \textsc{Togedule}. While the LLM component in \textsc{Togedule} effectively handles the complexity of representing group availability, its decision-making process lacks transparency. To enhance explainability, we retrieve reasons from GPT-4, as illustrated in Table~\ref{tab:gpt-examples}, evaluating how many time options to show to the next participant. We also observed a correlation between GPT-4's decisions and other metrics, such as the number of promising times or response rate, as shown in Figure~\ref{fig:llm-decisions}. Despite this, the specific mechanisms behind GPT-4's decision-making remain as a black box. Future work can further enhance the interpretability of LLM decisions in coordinating adaptive representations.

\subsection{Tradeoff between reducing the information load and incorporating complexities}

Although \textsc{Togedule} managed to enhance some of the task load indices on average, it also worsened other types of task load such as effort level or perceived performance. This may be because \textsc{Togedule} tried to incorporate various complexities. \textsc{Togedule} provides attendees a way to indicate three different levels of preferences. It also gives the user a choice to switch the view between shared calendar and poll to see more or fewer options. Furthermore, \textsc{Togedule} formalizes the priority of attendees by allowing organizers to input how important each attendee is. These features are helpful in that they enrich the information collected, but may deteriorate the task load of the users by showing them too much information and choices at once. In other words, there is a tradeoff between decreasing the information load and encoding the complex constraints to the system. Future work can further explore such a tradeoff to find a balance between the two and see if the issue of information load can be resolved as users learn how to use the system and become used to it.  

\subsection{The Role of Adaptive Representation}

We believe that the core idea of a dynamic scheduling tool adjusting the pool of time options in real-time can change the scheduling practice to be more efficient. In our experiments, narrowing the pool in response to the currently collected information was found to significantly reduce the need for verbal communication. This is a promising result that can inspire future work to further extend the dynamic scheduling tools that can possibly eliminate the need for cumbersome back-and-forth emails. This also aligns with the recent findings that early respondents play a vital role in driving the scheduling process to be successful and later respondents rarely mark availability for all the options presented.

It is also important to note that organizers using \textsc{Togedule} made decisions with higher quality and better speed. This suggests that scheduling tools should provide more support for organizers in deciding the optimal time rather than letting them manually track constraints such as priority or preference. In particular, as suggested in our formative study findings, our main study results also show that the decision support for organizers should be open-ended, giving more control over the recommendation to the organizers. Existing tools tend to provide a rigid recommendation. For instance, Google Calendar outputs a single set of best times to meet, and users are not able to see what happens if they change some of the constraints or use a different way to calculate the optimal time choice. \textsc{Togedule}’s improvement is that it shows a wider range of recommendations based on multiple possible algorithms and allows the organizers to interactively update the recommendations by changing the constraints such as the priority of each attendee. 

\subsection{Limitations and Future Work}

Our study has several limitations that need to be recognized. First, the evaluation of \textsc{Togedule} was conducted as a controlled experiment, where complexities were augmented under assumptions. \blue{While these experiments were carefully designed to simulate real-world scenarios, they may not fully capture the intricate dynamics and unpredictability of actual scheduling practices.} To address this, a deployment study in real-world environments would provide more valuable insights into the effectiveness of adaptive scheduling tools. Future work should explore this avenue to better understand how the system performs when confronted with authentic schedules, real constraints, and diverse user behaviors.

Second, we are not certain if the tradeoff between speed and quality in the current experiment design was representative of the real-world users. For instance, since we paid bonus payment not only for accuracy but also for speed, some organizers might have rushed in deciding the time to meet, which might have caused fewer decisions to be high quality across all conditions --- less than half of the submissions were considered good. Such cases might have resulted in the exaggerated effect of \textsc{Togedule} on the quality of decisions made by organizers. Future studies can try different incentive schemes to encourage the speed-quality tradeoff to resemble the real-world situation as much as possible.

Third, \textsc{Togedule} is still limited in that it does not \textit{fully} incorporate complexities around scheduling meetings. \textsc{Togedule} tries to take preference into account by allowing attendees to express three levels of preferences and priority by enabling organizers to set up the priority of each attendee. Still, \textsc{Togedule} is far from perfect and does not deal with \blue{various other} factors. Formalizing all possible complexities is prone to giving extra load for the users, so we chose to simplify some of the complexities rather than supporting even minor circumstances. Therefore, although \textsc{Togedule} further considers complexities that existing tools have not formalized, it still remains limited in some ways.

Despite such limitations, we believe that the \textsc{Togedule} system and findings from its evaluation have meaningful contributions to the CHI \blue{and CSCW} community moving forward. First, we propose a novel concept of dynamic scheduling that adjusts the options and their presentation format with regard to the currently collected information about the group’s availability. Second, the findings from our formative study show the need for dynamic scheduling tool and their potential benefits and challenges. Third, results from our experiments suggest that dynamic scheduling can be a promising way to reduce verbal communication and let organizers make high-quality decisions in less time. \blue{Lastly, \textsc{Togedule} demonstrates how AI can mediate collaboration, reducing the cognitive burdens associated with coordinating group availability. Its design has the potential to reshape group dynamics and decision-making processes by dynamically adjusting available time slots and their presentation format.}

\section{Conclusion}

Navigating and consolidating group availability for scheduling meetings poses a persistent challenge. Traditional tools often present a static array of options to all participants, irrespective of evolving inputs and individual preferences. In response, we introduce \textsc{Togedule}, a system that (1) dynamically utilizes prior inputs to optimize the set of choices presented to attendees and (2) recommends time slots for organizers to consider as the final meeting time using multiple algorithms with different weights and assumptions. We found that \textsc{Togedule} can alleviate cognitive burdens on attendees and help reduce the need for verbal communication. We also found that organizers using \textsc{Togedule} tend to make decisions with higher quality in a shorter amount of time.

\nocite{*}
\bibliographystyle{refs}
\bibliography{refs}


\end{document}